\pdfoutput=1
 
\documentclass[ reprint,
 amsmath,amssymb,
 aps,
]{revtex4-1}

\usepackage{color}
\usepackage{graphicx}% Include figure files
\usepackage{dcolumn}% Align table columns on decimal point
\usepackage{bm}% bold math
\begin{document}

%\preprint{APS/123-QED}

\title{A simple model clarifies the complicated relationships of complex networks}%

\author{Bojin Zheng}
\email{zhengbojin@gmail.com}
\affiliation{1 College of Computer Science, South-Central University For Nationalities, Wuhan 430074, China}
\affiliation{2 State Key Laboratory of Networking and Switching Technology, Beijing University of Posts and Telecommunications, Beijing 100876, China}
\affiliation{3 Computer School, Wuhan University, Wuhan 430072, China}

\author{Hongrun Wu}%
\affiliation{3 Computer School, Wuhan University, Wuhan 430072, China}

\author{Li Kuang}%
\affiliation{3 Computer School, Wuhan University, Wuhan 430072, China}

\author{Jun Qin}%
\affiliation{1 College of Computer Science, South-Central University For Nationalities, Wuhan 430074, China}

\author{Wenhua Du}%
\affiliation{1 College of Computer Science, South-Central University For Nationalities, Wuhan 430074, China}

\author{Jianmin Wang}%
\affiliation{4 School of Software, Tsinghua University, Beijing 100084, China}

\author{Deyi Li}%
\affiliation{4 School of Software, Tsinghua University, Beijing 100084, China}

\date{\today}% It is always \today, today,

\begin{abstract}
Real-world networks such as the Internet and WWW have many common traits. Until now, hundreds of models were proposed to characterize these traits for understanding the networks. Because different models used very different mechanisms, it is widely believed that these traits origin from different causes.
However, we find that a simple model based on optimisation can produce many traits,
including scale-free, small-world, ultra small-world, Delta-distribution, compact, fractal, regular
and random networks. Moreover, by revising the proposed model, the community-structure networks are generated. By this model and the revised versions, the complicated relationships of complex networks are illustrated.
The model brings a new universal perspective to the understanding of complex networks and provide a universal method to model complex networks from the viewpoint of optimisation.
\end{abstract}

%\keywords{complex network; origin; relationship}

\maketitle

%\tableofcontents

\section*{Introduction}
Complex networks have been found to be efficient and effective in illuminating various biological,
social, and technological systems \cite{34,80,308,310}, for examples, the Internet\cite{1,228}, WWW and protein-interaction networks\cite{300}. Through the efforts of many scientists,
numerous traits of complex networks, such as the scale-free property \cite{116},
the small-world effect\cite{16,305,306}, the community structure \cite{116}
and the fractal structure \cite{300,301}, have been discovered. Such traits are the foundation to model the real-world networks for understanding their origins and mechanisms.

To explain such traits, hundreds of models have been proposed. For example,
the Watts-Strogatz (WS) model \cite{16} illustrates the origin of the small-world effect
and demonstrates the relationships of small-world networks, random networks and regular networks:
i.e., small-world networks are an intermediate form between random networks and regular networks.
The Barab\'{a}si-Albert (BA) model \cite{1,117} demonstrates the scale-free property of networks,
and Amaral et al. \cite{302} clarified the relationship between scale-free networks and small-world networks.
Li et al. \cite{303} demonstrated the relationship between scale-free networks and random networks through
the locality hypothesis. Song et al. \cite{301,300} proposed a method to define fractal networks,
which involves the relationship between small-world networks and fractal networks.

Generally speaking, based on current knowledge, complex networks can be categorized into many types according to the traits,
such as random \cite{303,33}, regular, scale-free \cite{117}, small-world \cite{16,306},
ultra small-world \cite{305}, community-structure, compact \cite{152}, fractal, and Delta-distribution networks.
However, the relationships among these types of complex networks only have been partially explored.

Considering the number of the proposed models \cite{18,27,168,301,303,33,305} that explains the types of complex networks, it is reasonable to believe that these types of complex networks would have different causes: different types of complex networks originate from different origins and different mechanisms.
 However, when a network has multiple traits, multiple different mechanisms should be used to explain their corresponding traits; and there should be an assembling mechanism to combine these mechanisms of traits together. The combinatorics would make such a schema quite complicated, no matter that there are hundreds of different mechanisms for only one trait. People has to solve the competition of these mechanisms as well, if we take the Occam's Razor for granted.

Here, by using only three common measures, the degree of nodes, the degree of edges, and the average shortest path length, we implemented a simple model based on optimisation that can produce random, regular,
scale-free, small-world, ultra small-world, compact, fractal and Delta-distribution networks.
Moreover, with a slight revision, the model also can produce community-structure networks. Furthermore, all traits and their combinations can be explained by  revising the proposed model. These
results suggest that we can illustrate the relationships of various types of complex networks under the framework of optimisation, and bring a new perspective on understanding the real-world networks such as the Internet and WWW.

\section*{Results}

A network or graph is a set of nodes with edges.
Regarding the nodes, the degree is the primary measurement. As to the edges,
the concept of edge degree has been defined in various ways. To characterize
the holistic features of the entire network, the average shortest path length
is widely used \cite{311}. These three measures are the most commonly used measures in the study of complex networks.

It may appear that these measures have no bearing on the
resultant types of complex networks. However, our model shows that there is an intrinsic relationship among them.
The types are determined by three common measures.

\subsection*{The Model}

As mentioned above, the model requires a definition on the edge degree. Because the degree is the most commonly used measure of nodes, the
degree of an edge could be defined as a function of the degrees of the two nodes at its
ends. Here, the edge degree is defined as the product of the power function of the degrees of
two nodes at both ends (see Fig. 1).

\begin{figure}
\centerline{\includegraphics[width=2.60in,height=2.51in]{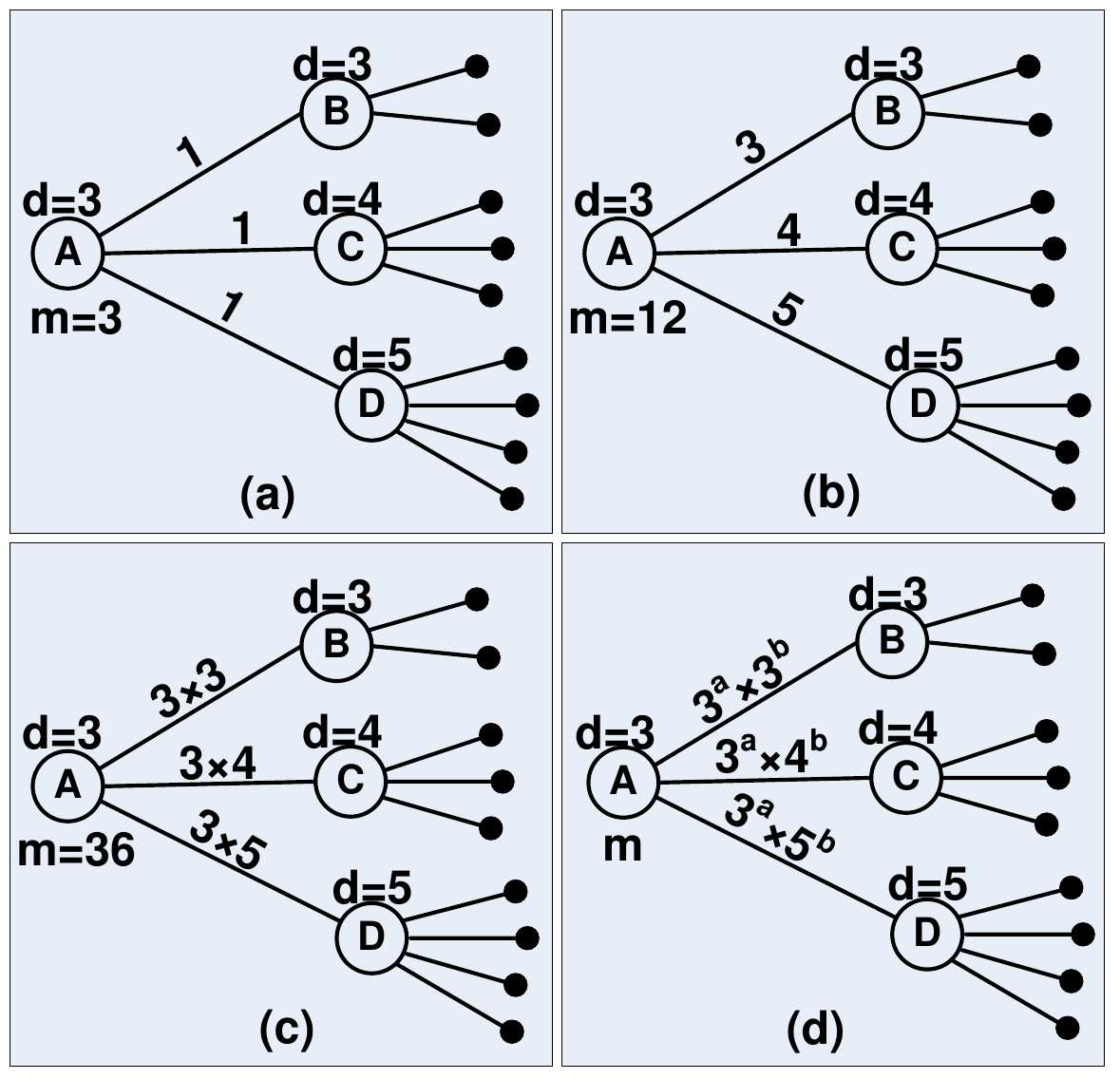}}
\caption{Definitions on the edge degree. (a) In the simplest case,
the edge degree of every edge is 1, irrelative to the degrees of
both nodes at the ends of the edge. (b) The edge degrees of node A
are the degrees of the neighbors, irrelative to the degree of node A itself. Here, regarding the nodes on two ends of an edge,
the degrees of an identical edge relative to the different nodes are different. (c) The edge degrees are the product of the degrees of nodes on the ends. (d) In the general form, the edge degree is the product of the power functions of the degrees of both nodes at the ends. The previous cases
are special cases with different values for $a$ and $b$.}\label{fig1}
\end{figure}

Based on the definitions above, the proposed model can be stated as follows.

\textbf{A connected undirected network evolves to
minimise the summation of the degrees of the nodes and to maximise
the summation of the degrees of the edges with a constant average shortest path length.}

That is, every network is evolving and should be optimised to achieve two objectives with a constraint on its average shortest path length.

Mathematically, this model is expressed by equation (\ref{eq1}).

\begin{equation}
\label{eq1}
\begin{array}{l}
 \left\{ {_{\max F_2(A)  = \sum\limits_{i = 1}^N {(\sum\limits_{j = 1}^N {x_i^ax_{j}^b\delta _{ij} } })}^{{\rm{min }}F_{\rm{1}}(A)  = \sum\limits_{i = 1}^N {x_i} } } \right. \\
s.t. \\
 y = c \\
 N > x_i \ge {\rm{xmin}} \\
 \end{array}
\end{equation}

Here, $x_i$ is the degree of node $i$, $y$ for the average shortest path length, $A$ for the evolving network, and \textit{c/xmin/a/b/N} are non-negative constants.
Furthermore, \textit{xmin} is the minimum degree of the nodes throughout
the entire network. The function \textit{$\delta $}$_{ij}$ is equal to 1 when a link between node $i$ and node $j$ exists; or it equals 0.

In equation (1), the proposed model is a bi-objective optimisation problem. The proposed model has feasible solutions, each solution indicating a network, and every best solution is a desired resultant network.

As to single-objective optimisation problems, the concept of ``the best solution''
is easy to understand. If one solution has the largest
function value for a maximisation problem or the smallest function value for
a minimisation problem, then it is the best solution. However,
bi-objective optimisation problems are quite different \cite{65}. Commonly, the solution with the
best function value for the first objective is far from the best for the second objective.
Therefore, the concept of ``the best solution'' must be extended in bi-objective optimisation problems.

The simplest way to extend this concept is to define ``the best solution'' as ``no solution is better
at satisfying both objectives''. This extended concept often
results in multiple best solutions. Because none of the
best solutions are dominated by a feasible solution,
they form a non-dominant set, which is known as the ``Pareto front'', a term coined by
David E. Goldberg \cite{66} in honor of V. Pareto \cite{29}.
By the way, another great achievement
of V. Pareto is the finding of the power law phenomenon in the wealth distribution.
For more detailed information on the Pareto front, please refer to the SI.

For any given parameter setting, there is a Pareto front for the proposed model. When optimisation algorithms are used to solve the proposed model, they actually obtain sampling points of Pareto front. According to these sampling points, the resultant networks can be constructed.

With the implementation of different parameters, the obtained network would exhibit
different traits and would correspond to different types. Because theses types are obtained for
the same model, the origin of these types and the relationships of the types can be determined.

\subsection*{Types of networks}

Researchers have observed many types of complex networks. Here, we discuss the most
common types: i.e., the scale-free, small-world, ultra small-world, fractal,
community-structure, compact, Delta-distribution,
random, and regular networks. Here, we theoretically demonstrate
that these common complex networks can be produced by the model described above.

\subsubsection*{Scale-free network}

The most popular theoretical description of scale-free networks is the BA model \cite{1}.
However, if we treat the node degrees as a random variable,
the proposed model can also produce scale-free networks. Obviously, some scale-free networks that satisfy the equation (\ref{eq1}) are in the Pareto front, while others are not. Here, we demonstrate that
the proposed model can produce scale-free networks in the Pareto front,
which we refer to as optimal scale-free networks.

When discussing the scale-free property or and random networks, we actually are discussing the degree distribution, i.e., treat the degree values as samples of a random variable. Therefore, here we treat $x_i$ and $x_j$ as samples of the random variable $X$. Because the
samples are independent and identically distributed, based on the Lagrangian
relaxation method\cite{nlp}, equation (\ref{eq1}) can be rewritten as equation (\ref{eq3}).

\begin{equation}
\label{eq3}
\begin{array}{l}
 \left\{ {\begin{array}{l}
 \mbox{min }f_1 (x_i) = x_i\mbox{ + }\theta(y - c)^2 \\
 \min f_2 (x_i) = \left( {\sum\limits_{j = 1}^N {x_{i}^{a}x_{j}^{b}\delta _{ij} }
} \right)^{\mbox{ - 1}}\mbox{ + }\theta(y- c)^2 \\
 \end{array}} \right. \\
 s.t.\mbox{ }N > x_i \ge xmin \\
 \end{array}
\end{equation}

Here, \textit{$\theta $} is an arbitrary positive real number.

Because $x_i$ and $x_j$ come from the same random variable, we use $x_i$ to approximate $x_j$, so $f_{2}$ can be further rewritten as equation (\ref{eq4}).

\begin{equation}
\label{eq4}
f_2 (x_i) \cong x_i^{\mbox{ - (1 + a + b)}}\mbox{ + }\theta(y - c)^2
\end{equation}

Equation (\ref{eq4}) has an analytic solution of a Pareto front \cite{155}, which can be
rewritten as equation (\ref{eq5}), when $y=c$, where $c$ does not constraint the
random variable $X$ through the validation of the network topology structure.

\begin{equation}
\label{eq5}
f_2 (x_i) = (x_i)^{ - (1 + a + b)}
\end{equation}

Because $f_{2}$ is a function that can be defined on the sample space, we can obtain
 equation (\ref{eq6}).

\begin{equation}
\label{eq6}
p(X) = C(X)^{ - (1 + a + b)}
\end{equation}

Here, $C$ is a constant to normalise $p(X)$ and satisfies the equation (\ref{eq7}).

\begin{equation}
\label{eq7}
C = \frac{1}{\sum\limits_{X = \mbox{1}}^{N\mbox{ - 1}} {(X)^{ - (1 + a +
b)}} }
\end{equation}

Equation (\ref{eq6}) indicates that under the condition that \textit{a$ \ne $0} or \textit{b$ \ne $0} and when $c$
does not constraint the distribution of $X$, i.e., is proper, the network is scale-free,
 and the exponent of the degree distribution obeys equation (\ref{eq8}).

\begin{equation}
\label{eq8}
\gamma \mbox{ = 1 + }a + b
\end{equation}

According to the definition of the optimal scale-free network, all optimal scale-free networks are the best solutions of this model.

Regarding the non-optimal scale-free networks, when $F_1$ is fixed, $F_2$ is not optimal: i.e.,
the hub nodes are not linked together.
When the hub nodes are divided into two or more groups,
the network is called a community-structure network. Thus,
the non-optimal scale-free networks are actually community-structure networks
or transitional forms between optimal scale-free networks and community-structure networks.

\subsubsection*{Community-structure network}

Community-structure scale-free networks can also be depicted by this model
with a slight modification. With this modification, community-structure scale-free networks
become the best solutions of the new model.

Community-structure scale-free networks are non-optimal scale-free networks. Assume
that there are two identical communities linked by only one edge; when certain edges in no. 1 community
are moved to no. 2, $F_2$ of the entire network can increase as the average shortest path length decreases, and
simultaneously, no. 1 community loses some edges, resulting in an increased average shortest path length; that is,
we can reach a solution that exhibits a larger $F_2$ but with the same $c$.
Therefore, the community-structure scale-free networks are non-optimal.

To produce optimal community-structure scale-free networks, the proposed model should be modified.

In the real world, community structure often relates to similarity distances, such as geographic distances,
cultural distances or cognitive distances. By taking these distances into consideration,
optimal community-structure scale-free networks can be produced by an enhanced model (see the SI).
This result indicates the origin of the community-structure scale-free networks.

The modified model here can produce typical networks with community structures.
To address the other non-optimal scale-free networks, more constraints must be added. We leave these issues to future work.

\subsubsection*{Compact network and Delta-distribution network}

According to equation (\ref{eq1}), the average shortest path length of the network is a hard constraint, so the constant $c$ can alter the forms of the resultant networks. When $c$ does not constrain the forms of the networks, we say that $c$ is
proper.

A proper $c$ depends on the constant \textit{xmin}.
From equation (\ref{eq.powerlaw}), which is the continuous version of the power law distribution,
when \textit{$\gamma $} is determined, the probability of $X$ depends on the constant \textit{xmin},
so the proper $c$ would decrease as \textit{xmin} increases.

\begin{equation}
\label{eq.powerlaw}
p(X) = \frac{\gamma - 1}{xmin}\left( {\frac{X}{xmin}} \right)^{ - \gamma }
\end{equation}

According to the definition of $F_2$, when some hub nodes link to other hub nodes, $F_2$ is maximised.
When $F_2$ is maximised, if $c$ is proper, and the hub nodes tend to link together,
the obtained networks would have a single center. Because hub nodes are the similar nodes to link together, the obtained network is hierarchical: i.e., the obtained network is onion-structure \cite{onion,178} alike or compact.
In such networks, the hub nodes tend to form an interconnected core, and the non-hub nodes with similar degree link together and encircle the core hierarchically. Moreover, the lower the degree of the node, the farther the node stay from the center.

When $c$ decreases to force the degree distribution away from that of
a scale-free network, the hub nodes collect more edges until
the network finally becomes a star-like or Delta-distribution network.

\subsubsection*{Fractal network}

Scale-free networks have a degree distribution of the form $p(k) \sim k^{-\gamma}$.
According to the definition of self-similarity (i.e., when an entire object is exactly or approximately similar to a part of itself), scale-free networks can be regarded
as self-similar with respect to the probability of the degree or can exhibit a probabilistic similarity
when we treat $p(k)$ as a function.

Alternatively, Song et al. proposed a definition on fractality of complex networks over the length. In the box covering method, if the box number $N_B$ has a power law relationship with the maximum box diameter $l_B$, as shown in Equation (\ref{eq.fractality}), then the networks present fractality or similarity over different length scales. Here, the fractality actually is a type of structural similarity.

\begin{equation}
\label{eq.fractality}
{N_B} \sim {l_B}^{ - {d_B}}\
\end{equation}

Obviously, structural similarity over the length, which is expected  in a fractal network, is different to the definition of probabilistic similarity over node degrees.

Additionally, the diameter of the whole network is often positively relative to average shortest path length, hence a fractal network is often expected to exhibit
a power relationship between the node number and average shortest path length,
and this relationship is expressed in Equation (\ref{eq.factal.aspl}).

\begin{equation}
\label{eq.factal.aspl}
c \sim N^{\mbox{1 / w}} \ \ \ (w>1)
\end{equation}

Equation (\ref{eq.factal.aspl}) implies that the average shortest path length should be quite large.
In fact, because $c$ depends on $xmin$,
the average shortest path length of the network should change with $xmin$. When $xmin$ increases,
$c$ of the fractal network can be smaller than $ln(N)$ .
Here, the qualitative relationships of $N$, $xmin$, $c$ and $w$ require further investigation.

In the proposed model, because $c$ ranges from $1$ to $N-1$, the average shortest path length of the fractal network must be included.
When $c$ is in the ranges of the fractal networks, the scale-free networks should be stretched. That is, a larger value of $c$ forces some marginal nodes away from the center of network.
When applying the box covering method, the larger $c$, i.e., often the larger diameter, may result in a power law relation between the box number and the maximum box diameter possible, thereby result in structural similarity.

More detailed information and the simulation results on fractal networks are discussed in the SI.

\subsubsection*{Small-world network and ultra small-world network}
The small-world network exhibits a clear feature in which the average shortest path length is approximately $ln(N)$,
in addition to a larger clustering coefficient \cite{16}. The latter feature is easily satisfied.
Hence, we discuss the previous feature only.

According to the definition of the small-world property, when the average shortest path length of the obtained
network is given by $c \simeq ln(N)$,
the network is considered a small-world network.

Moreover, when $c \simeq ln(ln(N))$, the network is an ultra small-world network. For any given network, the number of nodes determined the maximum of degree values, i.e., the maximum of random variable $X$. According to equation (\ref{eq.powerlaw}), when $xmin$ increases, if we also increase the maximum of degree values, then we can keep the $\gamma$ fixed. The increase of $xmin$ and maximum of degree values means more edges in a network, and more edges means smaller average shortest path length, that is, the ultra small-world property could emerge under some circumstances.

\subsubsection*{Random network}

When $a=b=0$, $F_{2}$ reduces to $F_{1}$. Because $F_{1}$ should be minimised and
$F_{2}$ should be maximised, the minimisation of $F_{1}$ will completely violate the maximisation of $F_{2}$,
such that
every solution would belong to the Pareto front. Therefore, the resulting
networks are random if $c$ does not constraint the distribution of $X$. When $c$ is small and closes to 1,
the network approximates a Delta-distribution network. When $c$ is
large, some nodes are forced to depart away from the denser center such that
the degree distribution resembles the power law distribution, with the amplitude ranging across several magnitude.
These results may imply a desirable study on the randomness and Zipf's-law-like distribution \cite{94}.

\subsection*{The Simulation}

Having theoretically analysed the produced types of networks, we now discuss the simulation results.

To solve this bi-objective optimisation problem by computer simulations, we use multi-objective optimisation
algorithms. Because $F_1$ is discrete, the histogram method (see the SI) is a suitable approach for
transferring this problem to a single-objective optimisation problem,
that is, first fix $F_1$, and only optimise $F_{2}$. Furthermore, to solve $F_{2}$, we employ a
greedy strategy. That is, we randomly generate a network and then continue to randomly change an
edge and update the network to a better solution. That is, if the change leads to a better $F_{2}$ and more closely
approximates the average shortest path, then we accept the change; otherwise, we refuse the change.
Besides, the proposed algorithm can be used to generate complex networks with arbitrary traits or the combinations of traits. For more information, see the SI.

Based on the method described above, we obtained various networks using different parameters.
Because this optimisation algorithm is a random algorithm, we performed this algorithm ten
times to verify its robustness. All of the runs that used the same parameters
generated similar results; thus, only the results obtained from the first run are shown (Fig. 2). Because we only used the greedy strategy, the resultant networks are local optimal solutions, not global optimal solutions. Although heuristic algorithms such as the simulated annealing algorithm \cite{250} can obtain the global optimal solutions, the computation time would be longer. Therefore we used the greedy strategy to obtain satisfactory results.

According to the theoretical analysis, the exponents of the degree distributions of the obtained networks
depend on $a$ and $b$; therefore, we designed 3 classes of experiments, with with $a=0$ and $b=0$, $a=0$
and $b=1$, $a=1$ and $b=1$, respectively. Because $xmin$ is related to $c$, we designed 3 sub-classes
of experiments, with $xmin=1,2,3$ for each of the classes. For each subclass,
we investigated various values of $c$. To show the generated networks clearly, the number of nodes $N$ in the simulations is set as 300. Also the simulations with larger size, the number of nodes with 1500, 3483 and 18000, are reported in SI.

From the experimental results, we chose some typical results to report in the SI. Here,
we selected 6 typical networks with $\gamma=2$($a$=0, $b$=1); the parameters and results are reported in Table 1, and the resultant topology is shown in Fig. 2.
%
%\begin{table}[ht]
%\caption{The parameters and results of selected networks.$E$ is the fixed value of $F_1$,
%$\gamma'$ is the exponent of the obtained network, $y$ is the actual average shortest path of the obtained network.}
%\begin{tabular}{lccccc}
%\hline
%\textit{No.}& $E$ & $c$ & \textit{xmin}& $\gamma'$ & $y$\\
%\hline
%\textit{(a)}&762&3.9&2&2.10&3.9 \\
%\textit{(b)}&762&5.5&2&2.11&5.5 \\
%\textit{(c)}&762&7&2&2.13&7 \\
%\textit{(d)}&1157&3.1&3&2.16&3.1 \\
%\textit{(e)}&1157&4.5&3&2.19&4.5 \\
%\textit{(f)}&1157&5.0&3&2.28&5.0 \\
%\hline
%\end{tabular}
%\end{table}

\begin{figure*}[ht]
\centerline{\includegraphics[width=6.19in,height=3.67in]{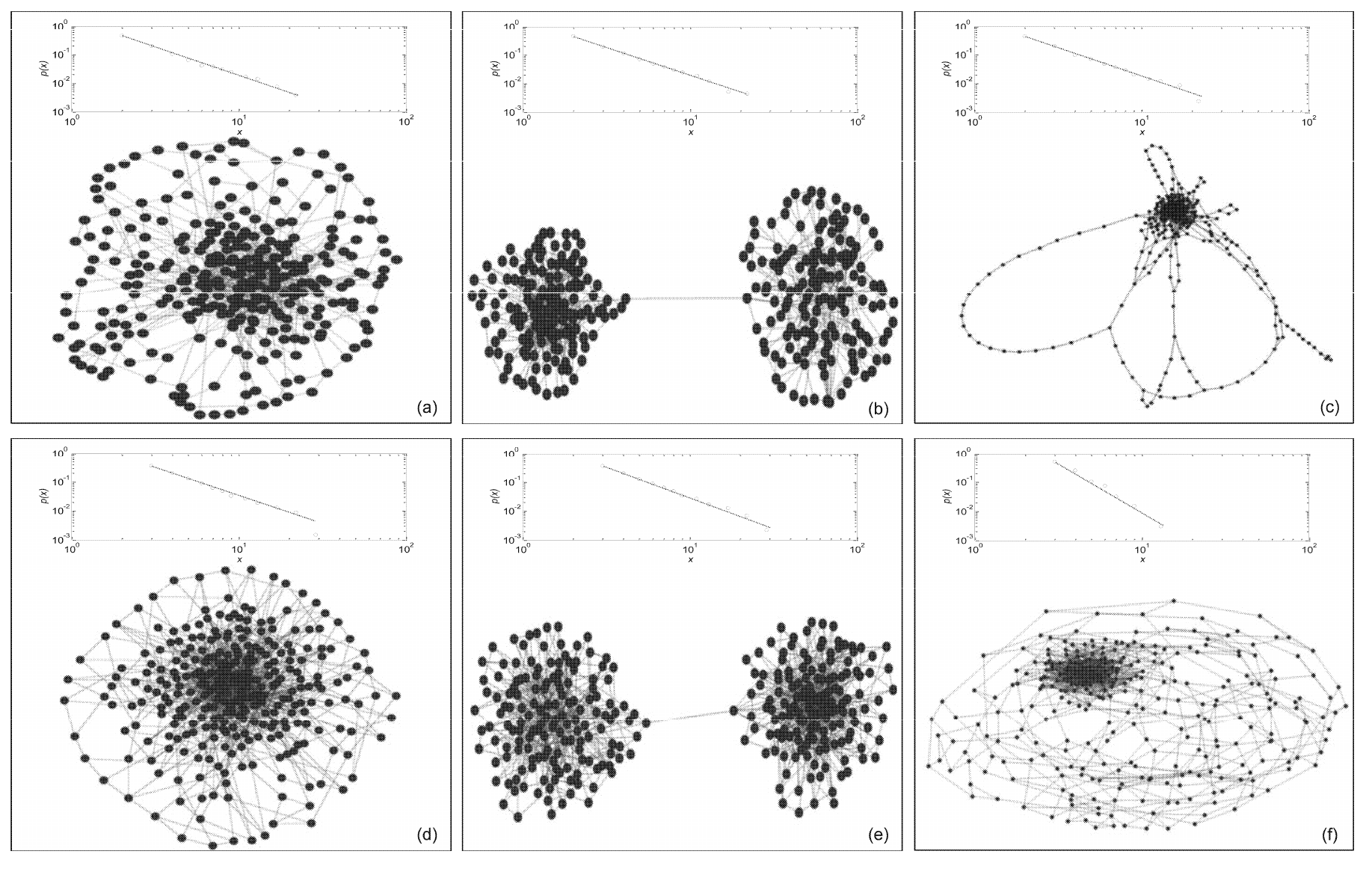}}
\caption{Typical networks and their degree distributions. The upper box
in each subfigure shows the degree distribution of the network in
the lower box. The degree distributions are plotted in a
log-log coordinate system. (a) This resultant network is
a compact network, whose \textit{c} is smaller than $ln(N)$. (b) This resultant network
demonstrates a network with two equivalent communities. (c) This beautiful network
is a fractal network. (d) This resultant network is also a compact network but
with denser edges. (e) This resultant network is a community-structure network.
Each community has denser edges. (f) This resultant network is a fractal network. The community-structure networks (b) and (e) are generated by the revised model in the SI, and the networks with multiple communities are shown in the SI; the fractality of (c) and (f) are also shown in the SI.
}\label{fig2}
\end{figure*}

Fig. 2 shows the compact, community-structure and fractal networks. The rows of the sub-figures show
the effect of $c$. When $c$ increases, the network type changes from compact to
fractal. The columns of the sub-figures show the effect of $xmin$. When $xmin$ increases, the
network average shortest path length for the same type decreases. Besides, we can see that the fractal networks here demonstrated the hub aggregation behaviors.

The results in Table 1 and Fig. 2 indicate that the obtained networks fit the power law distributions \cite{clauset2009power}. Besides, statistical evaluations on the fitness of the distribution of resultant networks are also reported in SI.
As shown in Table 1, the exponents of the networks are approximately equal to the expected values, and the
expected average shortest path length were also obtained.

Moreover, we observed that the community-structure networks exhibit a wide range of
values of $c$ because they can change the link(s) between the communities
to adapt to the topological distance. When $c$ is smaller, the link can connect the central
nodes of the communities; when $c$ is larger, the link can connect two marginal nodes in different communities.
For fractal networks, when $c$ reaches a certain value, the network is stretched.
As $c$ increases, the network first exhibits many circles and then becomes linear with a head that
exhibits dense nodes and edges.

In general, this model can generate various types of networks,
including small-world, ultra small-world, scale-free, community-structure,
and compact networks. Some types of the obtained networks are strongly dependent on the average shortest path length $c$ . However,
because there are no accurate definitions for the various types of networks, we cannot determine an accurate $c$
for each type from the experiments; we can only determine the relative relationships between
the types and the parameters. For more details on the results, please refer to the SI.

\section*{Discussion}

According to the simulation and theoretical results, the relationships of complex networks
can be illustrated under the framework of the proposed model.

Here, we assume that $N=300$, $\gamma=2$ and show a schematic map of the relationships in Fig. 3.
When $N$ or $\gamma$ changes, the schematic map also changes.

\begin{figure*}[ht]
\centerline{\includegraphics[width=0.618\linewidth]{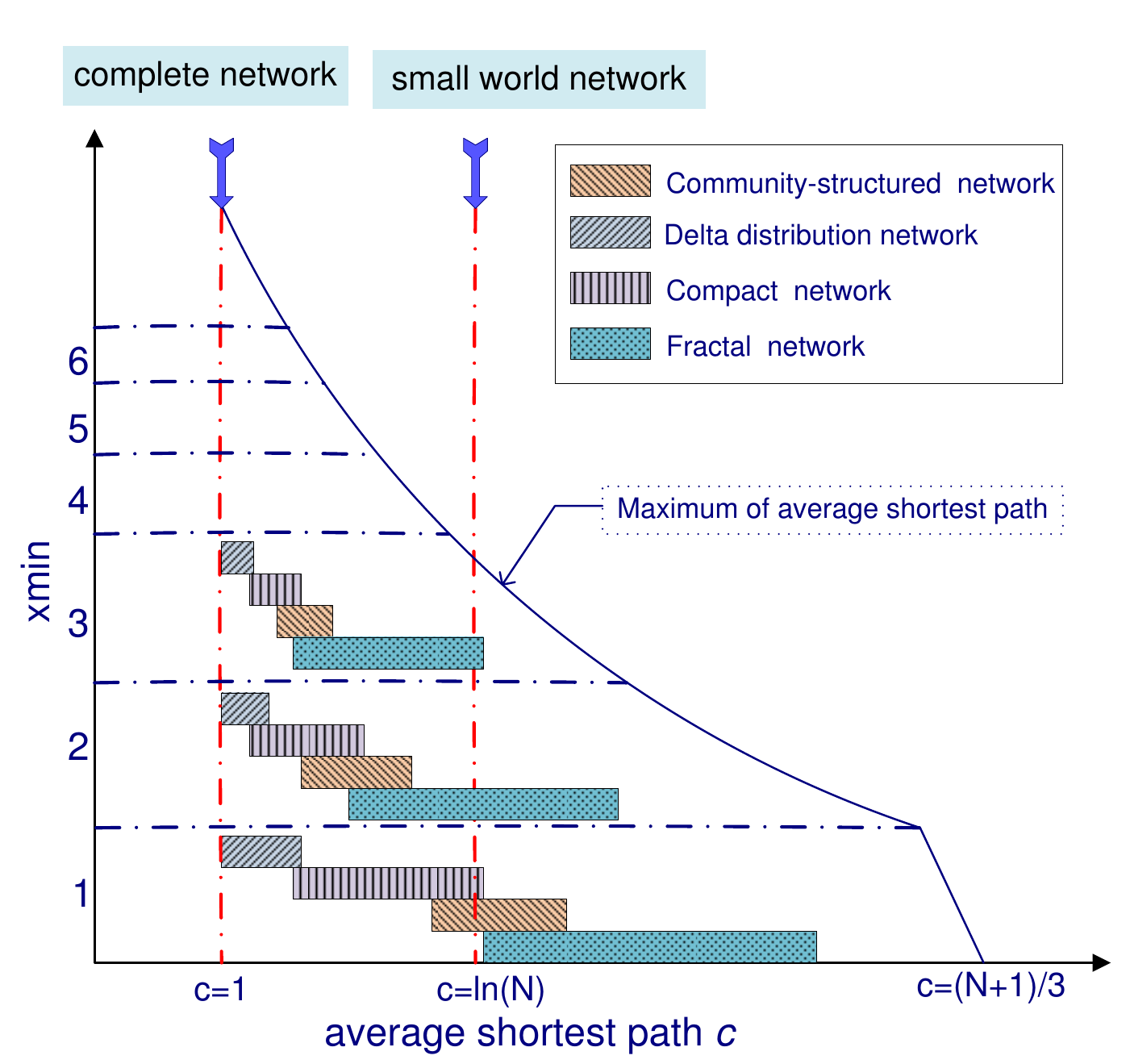}}
\caption{The schematic map on the relationships among various complex networks.
This figure assumes $\gamma$=2. When $\gamma$ varies, this figure would
also vary slightly. When $c$=1, the network is the complete network. When $c$=1, the generated network will be a complete network. With $xmin$=1, when $c$ increases starting from 1, firstly the resultant network is a delta-distribution network; when $c$ increases continuously, the resultant network is a compact network; when c increases continuously, the resultant network can be community-structure scale-free network if considering the similarity distance; when $c$ increases continuously, the resultant network is fractal network; when $c$ achieves the maximum, the resultant network is a linear regular network; when $c$=$ln(N)$, the resultant network is a small-world scale-free network. When $xmin$=2 and the other parameters keep the same, the order of the types of networks remains the same, but the spectral line(the positions of $c$) shift left and the ranges on $c$ decrease. For example, the generated network is small-world network when $xmin$=1 and $c$=$ln(N)$, but when $xmin$=3 and $c$=$ln(N)$, the network changes to be fractal network, and the result is shown as Fig.2(f). So when $xmin$ changes, the types also change.}
\label{fig3}
\end{figure*}

From Fig. 3, we can see that the average shortest path length can be regarded as a spectral line to discern the types of networks. With the increase of $c$, the order of the types is complete network, delta-distribution network, compact network, community-structure, fractal network. The other parameters, $xmin$ and $\gamma$ also affect the types of networks. When $xmin$ increase yet the other parameters keep the same, the sequence for the types of networks remains the same, but the spectral line shift left and the ranges of network types on $c$ decrease. The schematic map on $\gamma$=3 is shown in SI.

Based on the proposed model, the scale-free network plays a key and central role,
and scale-free networks can be categorized into several classes.
First, the scale-free networks can be divided into two types,
optimal and non-optimal. Optimal scale-free networks include the ultra
small-world, small-world, compact, and fractal networks,
which are controlled by the average shortest path length constraint. Outside of the optimal scale-free
networks but in the Pareto front, there are the Delta-distribution
and regular networks. Regarding the non-optimal scale-free networks,
there are community-structure networks and transitional forms between optimal
scale-free networks and community-structure networks. Moreover,
scale-free networks can be classified by an exponent. When the exponent is larger than 1,
the resulting networks are scale-free. However, when the exponent equals 1, the networks can be random.

In general, we demonstrated that a simple model can produce many common types of complex networks,
including scale-free, small-world, ultra small-world, community-structure, compact,
fractal, Delta-distribution, regular and random networks in this paper.
Our results indicate that three key measures can determine many types of complex networks.
Moreover, because these types originate from the same model,
their relationships can be illustrated under the framework of the proposed model.

%Considering that complex networks are often regarded as models of complex systems,
The proposed model brings a new perspective for understanding the complex networks and a new paradigm for distinguishing the
explanations of origins and mechanisms. When the proposed model is used to describe a certain complex network, it provides only one explanation on the origin and leaves the explanations of the mechanisms to the optimisation algorithms.
For instance, if we use a genetic algorithm to solve the proposed model,
then the genetic mechanism (or evolutionary mechanism) can be regarded as the mechanism of
the modeled complex network. That is, the mechanisms of complex networks
can be diverse while still representing similar phenomena.

%Another new perspective involves the physical meanings provided by the proposed model when
%applied to complex systems. For example, if this model is used to study the scale-free property of
%the Internet or food webs, the meanings of $F_1$ and $F_2$ would be very interesting. As to complex networks,
%$F_1$ and $F_2$ can be regarded as the improvement of invulnerability with the maximal efficiency \cite{152}.

Besides, physicists have used the optimisation to explain the world for centuries, for examples, the Fermat principle and the principle of minimum free energy etc.. Here our model is another example. By the optimisation method, we can characterize all the traits and their combinations, so the optimisation provides a universal method to model the real-world networks such as the Internet, WWW and protein-interaction networks. The ideal modeling networks generated by this universal method are useful of exploring the dynamics on complex networks, such as the synchronization, epidemic spreading and gaming.

\section*{Methods}

This paper first proposed an optimisation model based on three commonly used measures, i.e., the node degree, the edge degree and the average shortest path length. To solve this optimisation model, an algorithm with the greedy strategy was proposed. To obtain complex networks with larger sizes, a fast but specific algorithm was proposed. When solved this optimisation model, complex networks with different traits were obtained. According to the parameter settings of the proposed model, the relationships of traits of complex networks were illustrated. The details please refer to the SI.

%% == end of paper:

%% Optional Materials and Methods Section
%% The Materials and Methods section header will be added automatically.

%% Enter any subheads and the Materials and Methods text below.
%\begin{materials}
%\end{materials}

%\begin{materials}
%\section{AAA}
%\end{materials}

%\bibliographystyle{nature}
%\bibliography{ComplexNetwork}

\begin{table}[ht]
\caption{The parameters and results of selected networks.$E$ is the fixed value of $F_1$,
$\gamma'$ is the exponent of the obtained network, $y$ is the actual average shortest path of the obtained network.}
\begin{tabular}{lccccc}
\hline
\textit{No.}& $E$ & $c$ & \textit{xmin}& $\gamma'$ & $y$\\
\hline
\textit{(a)}&762&3.9&2&2.10&3.9 \\
\textit{(b)}&762&5.5&2&2.11&5.5 \\
\textit{(c)}&762&7&2&2.13&7 \\
\textit{(d)}&1157&3.1&3&2.16&3.1 \\
\textit{(e)}&1157&4.5&3&2.19&4.5 \\
\textit{(f)}&1157&5.0&3&2.28&5.0 \\
\hline
\end{tabular}
\end{table}

\section*{Acknowledgments}
We are grateful to Oskar Burger, Chunlai Zhou, Aimin Zhou, Baobin Wang, Weiwu Wang,
Yanni Han, Jun Hu, Yuanxiang Li, Guishen Chen, Haisu Zhang, Yutao Ma,
Jun'an Lu, Di Ning, and Xianjun Shen for many discussions, Shenzhan Li, Fei Xu, and Biao Wang
for their experimental assistance, and Yang Yang, Alan C. and Kristi H. for language assistance in writing this paper.
B.Z. thanks the National Basic Research Program of China (No.
2014CB340401) and the State Key Laboratory of Networking and Switching Technology
(No. SKLNST-2010-1-04) and the State Key Laboratory of Software Engineering (No.
SKLSE2012-09-15) and the China Scholarship Council for the supports. D.L. thanks the
National Natural Science Foundation of China (No. 61273213 and 61272111) for the
supports. J.Q. is grateful for support from the Fundamental Research Funds for the Central
Universities (No. CZY12032 and CZY13010).

\section*{Author contributions}
B.Z. designed research; B.Z. and H.W. performed research; B.Z., H.W., L.K. and W.D. analyzed data and performed simulations; B.Z., J.Q., J.W. and D.L. wrote the manuscript; all authors discussed the results and reviewed the manuscript.

\section*{Additional Information}

\textbf{Supplementary information} accompanies this paper at http://www.nature.com/scientificreports

\textbf{Competing interests statement:} The authors declare that they have no competing financial interests.
%
%\textbf{License:} This work is licensed under a Creative Commons Attribution-NonCommercial-NoDerivs 3.0 Unported License. To view a copy of this license, visit http://creativecommons.org/licenses/by-nd-nd/3.0/
%
%
%\textbf{How to cite this article:}

\end{document}